\documentclass[modern]{aastex61}

\usepackage[normalem]{ulem}

\newcommand\kms{$\rm km\,s^{-1}$}
\newcommand\msun{$\rm M_\odot$}
\newcommand\lsun{$\rm L_\odot$}

\accepted{ApJ}

\shorttitle{ALMA super star cluster cloud in NGC 5253}
\shortauthors{Turner et al.}
\begin{document}

\title{ALMA detects CO(3--2)  within a super star cluster in NGC~5253}

\correspondingauthor{Jean L. Turner}
\email{turner@astro.ucla.edu}

\author[0000-0003-4625-2951]{Jean L. Turner}
\affil{UCLA Department of Physics and Astronomy, Los Angeles, CA 90095-1547, USA}

\author[0000-0002-0214-0491]{S. Michelle Consiglio}
\affil{UCLA Department of Physics and Astronomy,  Los Angeles, CA 90095-1547, USA}

\author[0000-0002-5770-8494]{Sara C. Beck}
\affiliation{School of Physics and Astronomy, Tel Aviv University, Ramat Aviv, Israel}

\author{W. M. Goss}
\affiliation{National Radio Astronomy Observatory, Soccorro, NM 87801 USA}

\author{Paul T. P. Ho}
\affiliation{Academia Sinica Astronomy and Astrophysics, Taipei, Taiwan}

\author{David S. Meier}
\affiliation{Department of Physics, New Mexico Institute of Mining and Technology,
 Socorro, NM 87801 USA}
 
 \author{Sergiy Silich}
 \affiliation{Instituto Nacional de Astrof\'isica, \'Optica y Electr\'onica, Puebla, M\'exico C. P. 72840}
 
 \author{Jun-Hui Zhao}
 \affiliation{Harvard-Smithsonian Center for Astrophysics, Cambridge, MA 02138 USA}

\begin{abstract}

We present  observations of  CO(3--2) and $^{13}$CO(3--2) emission near the supernebula in
the dwarf galaxy NGC~5253, which contains one of the best examples of a potential globular
cluster in formation.  The 0\farcs3 resolution images reveal an unusual
molecular cloud, ``Cloud D1", coincident with the radio-infrared supernebula. 
The $\sim 6$-pc diameter cloud
has a linewidth, $\Delta v = 21.7$~\kms, that reflects only the gravitational potential of the star
cluster residing within it.
The corresponding virial mass is $2.5\times 10^5$~\msun. The cluster appears to have
a top-heavy initial mass function, with $M_*\ga 1$-2~\msun. Cloud D1 is optically thin in 
CO(3--2) probably because the gas is hot.
Molecular gas mass is very uncertain but constitutes  $<$ 35\% of the dynamical mass
within the cloud boundaries. 
 In spite of the presence of an estimated
$\sim 1500-2000$~O~stars within the small cloud, the CO appears relatively undisturbed.
We propose that Cloud D1 consists of molecular clumps or cores, possibly 
star-forming, orbiting with more
evolved stars in the core of the giant cluster. 

\end{abstract}

\keywords{galaxies: star formation --- 
galaxies: star clusters: individual (NGC~5253 supernebula) --- HII regions --- ISM: molecules}

\section{Introduction} \label{sec:intro}

Giant young star clusters, with masses $>10^5$~\msun, contain thousands of massive stars 
within the space of only a few parsecs. Given the rapid rate of evolution of O stars, how other stars
can form in their presence to build a large star cluster remains an outstanding problem.
The closest young massive clusters are in nearby galaxies; at 
these distances subarcsecond resolution is required 
to study star formation on cluster scales. The Atacama Large
Millimeter/Submillimeter Array (ALMA) can now provide images of gas  at subarcsecond resolution
corresponding to cluster scales in local galaxies. 
The J=3--2 line of CO is bright and easily excited in  dense 
($n\ga 2\times 10^4~\rm cm^{-3}$) gas, and can be used to estimate 
gas masses and kinematics to study the star formation 
process and how feedback occurs within massive clusters.

NGC 5253 is a local (3.8 Mpc: 1\arcsec = 18.4 pc) dwarf spheroidal galaxy 
\citep{1989ApJ...338..789C,1998ApJ...506..222M}  
with many young star clusters
\citep{1989ApJ...338..789C,1995AJ....110.2665M,1996AJ....112.1886G,1997AJ....114.1834C,
2004ApJ...603..503H,2001ApJ...555..322T,
2013MNRAS.431.2917D,2015ApJ...811...75C} and an infrared luminosity of $\sim 10^9$~\lsun\ 
\citep{2004A&A...415..509V,2005A&A...434..849H}. Its  stellar mass is
$\sim 2\times 10^8$~\msun\ \citep[][the dark matter mass could be  ten
times larger]{1998ApJ...506..222M}.
At least one-third of the galaxy's infrared luminosity originates from a giant star-forming region,
 a compact ($\la 3$~pc) radio source known as the ``supernebula"
\citep{1996ApJ...457..610B,1998AJ....116.1212T,2001ApJ...554L..29G,2004ApJ...602L..85T,2007ApJ...670..295R}. Extinction toward the supernebula is high, $A_V \sim 16$-18
\citep{1997AJ....114.1834C,2003Natur.423..621T,2005A&A...429..449M,2015ApJ...811...75C}.
 The  supernebula is coincident with a bright 
infrared source \citep{2001ApJ...554L..29G,2003Natur.423..621T,2004ApJ...612..222A} 
that may coincide with a visible red cluster, \# 11 
\citep{2015ApJ...811...75C, 2016ApJ...823...38S}.
Based on CO observations, there appears to be little molecular gas
within NGC~5253; most of the CO emission is found in a streamer along the prominent minor axis dust lane 
\citep{1997ApJ...474L..11T,2002AJ....124..877M,2015PASJ...67L...1M}.
A bright CO(3--2) source was detected near the supernebula in 4\arcsec\ observations
with the Submillimeter Array; comparison to lower J CO images 
suggest very warm gas, $T\ga 200-300$~K, in the central regions \citep{2015Natur.519..331T}.

We present ALMA observations of  CO and $^{13}$CO J=3--2 emission 
at $\sim$0\farcs3 resolution (5.5 pc) within the central
region of NGC 5253. Requiring gas of density $>20,000~\rm cm^{-3}$ for collisional excitation, 
CO(3--2) traces the dense gas typically associated with star forming cores 
\citep[e.g.,][]{1985prpl.conf...81M,2010ApJ...724..687L}. 
The observations reveal a number of
dense clouds within the central $\sim 100$ pc starburst region identified as Cloud D by
the CO(2-1) analysis of \citet{2002AJ....124..877M}; one cloud stands out in terms
of its unusual properties. 
We present here 
an analysis of the molecular cloud coincident with the supernebula, henceforth denoted ``Cloud D1".  
\clearpage

\section{Observations} \label{sec:obs}

NGC 5253 was observed in Band 7, a Cycle 1 program (ID = 2012.1.00105.S, PI = J. Turner) executed in Cycle 2 on 2015 June 4 and 5. The pointing shown here is centered at 13:39:44.911910, -31:38:26.49657 (J2000). 
The full mosaic of NGC~5253 including $^{13}$CO(3--2) is presented elsewhere \citep{SMC17}.  Spectral windows have a total bandwidth of 937.500 MHz, with  244.141 kHz per channel. Velocities are barycentric, in radio convention.  Bandpass and phase were calibrated with J1427-42064 and J1342-2900 respectively. Titan was the flux calibrator.  Calibration was done with CASA, 
pipeline 4.2.2 by the Joint ALMA Observatory. Absolute flux calibration is to within
10\% for Cycle 2 Band 7 data \citep{L2013}. 
Imaging  was done with CASA pipeline 4.5.0 by the authors. The synthesized beam for the CO(3-2) maps is 0.33\arcsec\ $\times$ 0.27\arcsec\, p.a. -90\degr. The conversion to brightness is 1K $\sim$ 9-17 mJy, with the smaller value for point sources, and the larger for sources filling the beam. A continuum map was constructed from offline channels in the band and subtracted from the (u,v) data by the authors before making line maps. The shortest baselines in the image are 25-100 k$\lambda$; emission more extended than $\sim$4\arcsec\ ($\sim$75 pc) can be poorly represented in these maps; from comparison with SMA data we estimate that $\sim$50\% of the emission is in such faint, extended structure
 \citep{SMC17}. 
The RMS noise in the individual 1 \kms\ line maps is 2.7 mJy/bm for CO(3-2) at 345.796 GHz. The integrated intensity map  (Moment 0 map) was made by summing emission greater than $\pm 2.5\sigma$ in the  cube.

\section{Cloud D1 and the supernebula} \label{sec:results}

\begin{figure}[ht!]
\begin{center}
\includegraphics[angle=0,width=5in]{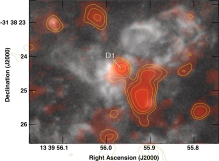}
\end{center}
\caption{ALMA CO J=3-2 emission in NGC 5253, in color and contours, superimposed on 
an archival HST H$\alpha$ image. The ALMA  beam is
0\farcs33 $\times$ 0\farcs27, p.a. -90.0\degr\ (6 pc $\times$ 5 pc).
 The larger CO cloud to the southwest of D1 is $\sim$15-20 pc away and redshifted with respect to D1. 
 Cloud D1 is the focus of this paper;  other clouds  are discussed 
in Consiglio et al.\ (2017). For registration with the HST image we assumed that D1, which is coincident
with the radio continuum nebula, 
 is coincident with the Pa $\beta$ source in Calzetti et al. (2015). }
\end{figure}

The ALMA image of CO(3-2) integrated line emission, 
in red and contours, 
is shown overlaid 
on an HST H$\alpha$ image in Figure 1. At ten times higher spatial resolution than previous CO maps, the ALMA CO(3--2) image reveals that what was previously identified as Cloud D \citep{2002AJ....124..877M,2015Natur.519..331T} actually comprises many molecular clouds. ``Cloud D1"
a compact and bright CO(3--2) source labeled in the figure, 
 is coincident with the core of the radio/infrared supernebula. 
 Fifteen to twenty parsecs to the southwest of D1 is a separate extended cloud with stronger CO(3--2) emission that is redshifted with respect to Cloud D1. There are numerous other clouds within the central region that are discussed elsewhere \citep{SMC17}. In this paper we focus on the unusual molecular Cloud D1. 

NGC~5253 is a galaxy  
 known for bright nebular emission as well as for the presence
of dust \citep{1962ApJ...135..694B,1970ApJ...159L.165K}. The radio 
and mid-infrared emission in NGC 5253 is dominated by the supernebula, a 
 source of radius $< 2$~pc
\citep[][]{1998AJ....116.1212T,2000ApJ...532L.109T,2001ApJ...554L..29G,2004ApJ...602L..85T}  located
near luminous and  young star clusters \citep{1995AJ....110.2665M,1997AJ....114.1834C}. 
From radio continuum fluxes \citep{1998AJ....116.1212T,2000ApJ...532L.109T,2004ApJ...602L..85T,2002AJ....124..877M,
 2007ApJ...670..295R}, and radio recombination line fluxes \citep{2007ApJ...670..295R,Bendo17}, 
 a Lyman continuum rate of  $N_{Lyc} \ga 3.3 \times 10^{52}~\rm s^{-1}$  is indicated
  \citep[corrected for direct dust absorption of UV photons following Inoue (2001),][]{2015Natur.519..331T}.  $N_{Lyc}$ is computed for an ionization-bounded nebula, and will be larger
   if there is photon leakage. Starburst99 models, described below, indicate that 
 this Lyman continuum rate corresponds to $\sim$1400-1800 O stars. 
The supernebula lies close to objects identified as 
Clusters 5 and 11 by \citet{2015ApJ...811...75C}; from their colors
 both appear to contain stars $\sim$1 Myr in age \citep{2015ApJ...811...75C,2016ApJ...823...38S}.

Cloud D1 is located within 0.6 pc in projection of the embedded 
supernebula \citep[Table 1;][]{2004ApJ...602L..85T}
The CO(3-2) line centroid,  v$_{bary}$ = 387.6$\pm$0.5~\kms, agrees to $<$2 \kms\ with the 
main 10.5$\mu$m [SIV] line from the ionized gas \citep{2012ApJ...755...59B} 
and to $<$3 \kms\ 
with the H30$\alpha$ radio recombination line \citep{Bendo17}
from the supernebula. Cloud D1 is 
thus coincident with the supernebula both in projection and in velocity, and it is similar if not 
identical in size.

\begin{deluxetable*}{lc}[b!]
\tablecaption{Cloud D1 in NGC 5253\label{tab:table}}
\tablecolumns{2}
\tablenum{1}
\tablewidth{0pt}
\tablehead{
\colhead{Quantity} & \colhead{Value}  \\
}
\startdata
RA(2000) & 13$^h$ 39$^m$ 55\fs9651$\pm$0\fs0004 \\
Dec (2000) & -31\degr 38\arcmin 24\farcs364$\pm$0\farcs006 \\
Assumed distance & 3.8~Mpc\\
V(CO3-2)\tablenotemark{a} &$ 387.6\pm0.5~\rm km\,s^{-1}$ \\
$\Delta$V(CO(3--2))\tablenotemark{b} & $21.7\pm0.5$ ~$\rm km\,s^{-1}$ \\
S(CO 3--2)\tablenotemark{c} & $2.2\pm0.2$~Jy $\rm km\,s^{-1}$\\
S($^{13}$CO3--2)\tablenotemark{c} & $0.05\pm0.01$~Jy $\rm km\,s^{-1}$\\
Radius\tablenotemark{d} & 0\farcs3 $\pm$ 0\farcs05 \\
$M_{vir}$\tablenotemark{e} & $2.5\pm0.9\times10^5$~$\rm M_\odot$ \\
$\rm N_{Lyc}$\tablenotemark{f} & $3.3 \pm 0.3 \times 10^{52}~\rm s^{-1}$\\
\enddata
\tablenotetext{a}{Line centroid, barycentric, radio definition.}
\tablenotetext{b}{FWHM; Gaussian fit to line in CASA.}
\tablenotetext{c}{Integrated line fluxes are for 0\farcs5 aperture centered on D1. Uncertainty in the
CO(3--2) flux is dominated by 10\% calibration uncertainty;  the $^{13}$CO(3--2) flux uncertainty  is 
due to signal-to-noise. See Observations.}
\tablenotetext{d}{Deconvolved from beam in CASA assuming a Gaussian source profile. See text.}
\tablenotetext{e}{Virial mass based on CO linewidth and size. 
For $\rho\propto 1/r$; for $1/r^2$ the mass is 30\% less.}
\tablenotetext{f}{Lyman continuum rate from the literature, primarily \citep{1998AJ....116.1212T,Bendo17},
and corrected for 30\% direct absorption of uv photons by dust \citep{2015Natur.519..331T} following the
procedure of 
\citet{2001AJ....122.1788I}. Corrected to 3.8 Mpc. Assumes ionization-bounded nebula and as such is a lower
limit to the true rate.}

\end{deluxetable*}

Cloud D1 is small. A Gaussian fit to the integrated intensity image (Fig.~1) gives a  size of
218$\pm$33mas $\times$ 104$\pm$54mas, p.a. 2.2\degr$\pm$77\degr, FWHM, deconvolved from the beam.
Fits of D1 within the 24 individual channel maps within the FWHM  give sizes of $\la$0\farcs3$\pm$
0\farcs05. 
We adopt a  size of 0\farcs3 $\pm$ 0\farcs05, or r=2.8~pc.
Excitation of CO(3-2) requires minimum molecular gas densities of $n\ga 20,000\rm ~cm^{-3}$, 
 so the molecular gas in Cloud D1 is dense. 

Cloud D1 is optically thin in 
the CO(3--2) line. 
 In Figure 2, 
 spectra are plotted for both  CO(3-2) and $^{13}$CO(3-2) lines within
  a 0\farcs5 region centered on Cloud D1. The integrated intensity of 
$^{12}$CO(3--2) is $\rm S_{CO}= 2.2\pm 0.2~\rm Jy\,km\,s^{-1}; $ for
$^{13}$CO(3--2)  is  $\rm S_{13CO}\sim 40$-$50\pm 8~\rm mJy\,km\,s^{-1}. $ 
The line ratio of $\sim$50 
for Cloud D1 is  close
to the abundance ratio of [CO]/[$^{13}$CO]$\sim$70 for the solar neighborhood,
and the [C]/[$^{13}$C]$\sim$40-50 estimated for the LMC  \citep[][]{1999AA...344..817H}. 
The Cloud D1 ratio is 
significantly higher than the $\sim11$-$15$ observed in the nearby clouds, including
the large cloud to the southwest of D1 \citep{SMC17}, which is 
closer to the value of 13 observed in the Orion molecular cloud in optically thick, cool 
CO gas \citep{1997ApJS..108..301S}. 
Warm gas can produce low optical depths in CO
and this is what we propose for D1.  For 300~K, as estimated from modeling
of the CO(3-2)/CO(1-0) line ratio \citep{2015Natur.519..331T},  the CO partition function
is $\sim$100, and the populations of the low J levels are correspondingly low. 
Thus the low optical depth in the CO(3-2)  line
from Cloud D1 is additional evidence that
the gas is in close proximity to the super star cluster.

A dynamical mass within  Cloud D1 can be obtained from the 
width of the CO line, 
$\Delta v= 21.7\pm 0.5$~\kms\ (FWHM). For an $r^{-1}$  mass profile, 
as in the Galactic cloud W49N, 
$M_{vir}$(r$<$2.8~pc)$\sim2.5\pm 0.9 \times 10^5 ~\rm M_\odot$. (An $r^{-2}$
profile giving $\sim$30\% less mass is included in the uncertainty.) This mass estimate assumes
a virialized cloud; it is unclear if this is valid for such a dynamic young source. 
However, most other
sources of velocity, such as protostellar outflows or stellar winds, 
 would if anything  broaden the line over the gravitational value. Moreover the mass
is close to that expected from the star cluster based on its radio free-free flux. 
Hence we adopt a gravitational mass of 
$2.5\pm 0.9 \times 10^5 ~\rm M_\odot$. The 
mean mass density within the cloud 
region, most of which is in stars, 
 is then 2700 \msun~pc$^{-3}$, corresponding
to  $<$$n_{H2}$$>$$\sim 40,000~\rm cm^{-3}$, and 
  surface density 
$\Sigma_{D1}=10^4~\rm \rm M_\odot ~cm^{-2}=~2.1~ g~cm^{-2}$,
 in the range observed for super star clusters \citep{2014prpl.conf..149T}.  
 These would
have been the original gas values at the onset of star formation in the cloud.

It is exceedingly difficult to determine a molecular gas
mass for Cloud D1. The metallicity of the region is unclear. The galaxy
overall is metal-poor, $Z\sim 0.25Z_\sun$, but
 there is evidence for localized enrichment near the young clusters 
\citep{1989MNRAS.239..297W,1997ApJ...481L..75S,1997ApJ...477..679K,2007ApJ...656..168L,
2015Natur.519..331T}.
From the CO(3-2) line intensity of $2.2\pm 0.2~\rm Jy\,km\,s^{-1}$,
and for $T_{ex} = 300$~K \citep{2015Natur.519..331T}, an uncertain number, 
the mass in CO molecules alone is $M_{CO}=3.0 ~\rm M_\odot$ for optically thin emission; then 
for a Galactic [CO]/[H$_2$] the molecular gas mass is  
$M_{H2} \sim 3500~M_\odot$ (includes He; for T=50K, the mass would be $\sim$4 times lower.)
 However, the Galactic [CO]/[H$_2$] abundance may not be appropriate in this environment. 
An alternate method using an empirical CO ``conversion factor", 
$X_{CO} = 4.7\times 10^{20}~\rm cm^{-2}
~(K~km/s^{-1})^{-1}$, gives $M_{gas}=6.5\times 10^4 ~\rm
M_\odot$ (including He). Since this  relation applies to gas-dominated, optically thick clouds, 
it should overestimate the gas mass in D1. The mass of ionized gas in the supernebula  is 
$M_{HII} \sim 2000$~\msun\ \citep{2004ApJ...602L..85T}; 
the estimated HI mass within D1 based on the  21 cm absorption column
\citep{2008AJ....135..527K} toward
the supernebula is minimal ($\sim$200-500~\msun; larger HI columns  will become molecular).
The combined molecular-plus-ionized gas mass within Cloud D1 is 
$M_{gas}\sim$6000-60,000~\msun. We estimate that gas constitutes $<35$\% of the mass within the boundaries of Cloud D1.
 
Based on the dynamical mass from the CO linewidth, which is dominated by stars, 
we can compare the mass, $M_{VIR}$, indicated for the cluster, 
and the observed Lyman continuum rate. 
We find $M_{dyn}/L\la0.0005$~\msun/\lsun.  
 We use Starburst99 models to model the cluster age and IMF. 
For the calculated  $N_{Lyc}^{corr}$ and $M_{dyn}$ (Table 1),  Starburst99 
\citep{1999ApJS..123....3L,2014ApJS..212...14L} models were run 
for Padova and Geneva models for metallicities of
Z=0.004 (closest to the mean metallicity of NGC~5253)
and 0.008 with Kroupa initial mass functions (IMFs) with exponent 2.3.  Statistical evidence
\citep{2004MNRAS.348..187W,2005ApJ...620L..43O,2013pss5.book..115K} suggests
an upper mass limit of $\sim$150~\msun\
for stars. Starburst99 models with an upper IMF
mass cutoff of 150~\msun\ require a lower IMF mass cutoff  $\ga$2 \msun\  to 
reach the observed $N_{Lyc}$ for a cluster  of $2\times10^5$~\msun. 
It has been suggested that stars of 
 200 \msun\ or more may exist in R136 and NGC~5253 
 \citep{2010MNRAS.408..731C,2016ApJ...823...38S}, perhaps
 formed from binary mergers \citep{2012MNRAS.426.1416B}. Such supermassive stars
 could increase the Lyman continuum rate without adding much mass. Even 
for a 200 \msun\ cutoff, Starburst99 models indicate a lower mass cutoff  $\ga$1~\msun\ for 
the D1 cluster. 
 If there is leakage of uv photons beyond the supernebula, which would increase
  $N_{Lyc}$, the lower  mass cutoffs for the IMF are even higher. 
 Another explanation for the low mass-to-luminosity
 ratio may be interacting binary stars \citep{2016MNRAS.456..485S}.

\section{The internal structure of Cloud D1}\label{sec:internal}

 Cloud D1 is located within the supernebula/cluster core, a harsh environment for molecules, where the mean separation between O stars is only $\sim$0.1 pc.
The CO  properties suggest that Cloud D1 is composed of many pockets of 
dense molecular gas, which may be in the form of
 protostellar disks, hot molecular cores surrounding individual
stars, or residual dense molecular clumps. 
 The estimated molecular gas mass of $M_{H_2} =3500$-$60,000~\rm M_\odot$, including He, 
 predicts a column density of 
 $N_{H_2} = 0.65$-$11\times 10^{22}~\rm cm^{-2}$, and 
 gas volume density of $M_{H2} = 570$-$9800~\rm cm^{-3}$. 
 Comparison with the CO(3--2) critical density of $\ga 20000~\rm cm^{-3}$ 
 indicates 
a volume filling factor of  $\la$3-50\%. We regard the lower value,
 corresponding to the optically thin mass, as more likely, and that the volume filling factor, 
 $f_{vol}\la 10\%$.

 \begin{figure}[ht!]
   \begin{center}
\includegraphics[angle=0,width=4in]{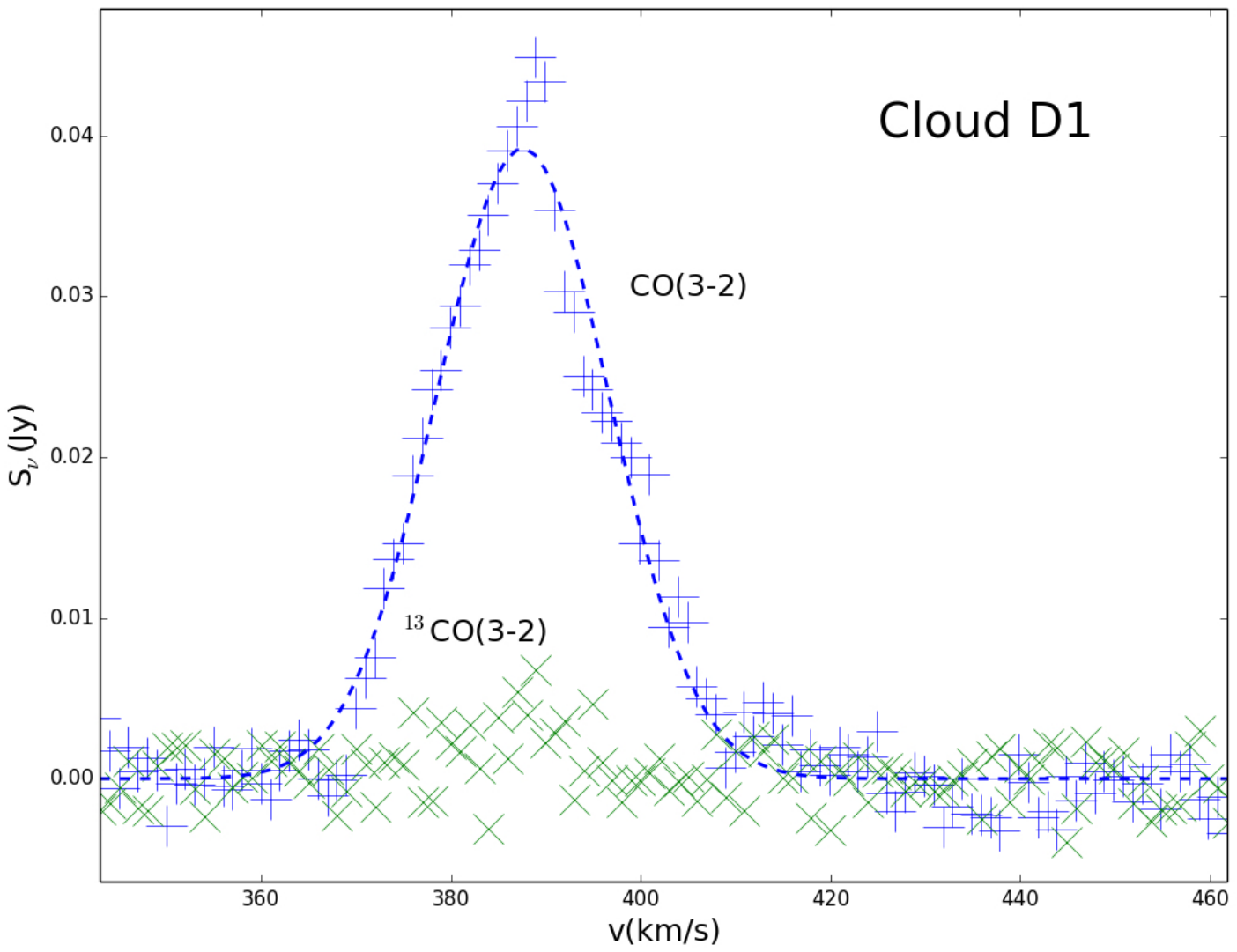}
\end{center}
\caption{CO(3-2) ($+$) and $^{13}$CO(3--2) ($\times$) line profiles of Cloud D1 in NGC 5253. 
Plotted are fluxes in individual channels of the inner 0\farcs5 circular region centered on the 
CO(3-2) integrated intensity peak in Cloud D1. Dashed curve is a least squares
 Gaussian fit to the CO(3--2) line. 
 Channels are 1 \kms. The rms is 2.7 mJy, approximately the size of the crosses. 
 Velocity is barycentric, radio convention. }
\end{figure}

The smooth and near-Gaussian CO line profile and the similarity in shape of D1 across the line
also suggests many clumps. Figure 2 shows the CO(3--2) line 
profile and the Gaussian fit. Departures from smoothness can be used to estimate numbers of clumps, even if the line is not perfectly Gaussian \citep{2008A&A...489..567B}. Channel-to-channel variations for the central 24 channels are typically  $\lesssim$20\%, implying $\sim$25 clumps per channel 
and $N_{cl}\gtrsim$600 for clumps with individual thermal linewidths of $\sim$1 \kms. 
 If instead Cloud D1 is composed of CO protostellar disks or molecular cores surrounding individual stars, with expected linewidths of 1-10 \kms, fewer disks/cores are needed to produce a smooth line. 
 Spatial variations in the centroid and deconvolved size
  of the emission from Cloud D1 across the 24 line channels are also consistent with no variation
   (centroids coincident to $\la$20 mas and in size, to $\la$50~mas), consistent with the many-clump hypothesis. 
   The slight nonGaussianity would be consistent with randomly distributed substructures. The blueshifted
side of the line is smoother than the redshifted side, which has no clear explanation; the redshifted
side may be related to the southeastern extension of the cloud. 

Are these small clumps or cores consistent with the radiation shielding necessary for the existence of CO?
There is a minimum size expected for CO-emitting clouds, since 
CO is chemically sustained only at $A_v>2$, or cloud column $A_v>4$ \citep{HT99,2015MNRAS.454.2828B}. 
At a density of $n_{crit}=30,000~\rm cm^{-3}$ (for T=300K), 
the minimum cloud column 
corresponds to a distance of 0.04 pc = 8000 AU, for a Galactic
$A_V/(N_H+2N_{H2})$ ratio. The corresponding minimum clump mass is 0.1~\msun. It is thus plausible 
that
Cloud D1 could  accommodate these minimum $A_v$ clumps both in linear dimension and in mass.

\section{Feedback and star formation within a young, massive cluster} \label{sec:feedback}

Cloud D1 coincides with the supernebula, which is an 
ultracompact HII region, and nearby evidence of Wolf-Rayet stars 
 \citep{1989MNRAS.239..297W,1997ApJ...481L..75S,1997ApJ...477..679K,2007ApJ...656..168L}
 that are typically 3-4 Myr in age, but sometimes less \citep{2016ApJ...823...38S}. Wolf-Rayet stars can lose copious amounts of metal-enriched mass. Thus the presence of molecular gas may not in itself give a good indication of cluster age.  
 Some super star clusters appear to be actively dispersing gas in winds, as appears
 to be occurring  in
 NGC 253 \citep{2006ApJ...636..685S,2013Natur.499..450B}. 
 However,
 if the cluster is sufficiently massive, theory suggests that the evolution of the HII region can be affected by gravity \citep{2002MNRAS.336.1188K, 2010ApJ...709..191M}, so that in some cases the enriched products of stellar mass loss may be retained by the cluster \citep{2017MNRAS.465.1375S}. Cloud D1 could perhaps survive after a supernova, since simulations suggest that gas can backfill into the cluster after the explosion \citep{2015ApJ...814L...8T}. We 
note that the CO linewidth reflecting the core cluster mass within the supernebula is almost precisely equal
to the thermal linewidth of the ionized hydrogen; this may not be coincidence.

To put Cloud D1 in perspective, it is instructive to compare to its closest Galactic analog, W49N. 
Of similar extent (r$\sim$3 pc) with ultracompact HII regions \citep{1984ApJ...283..632D,2000ApJ...540..308D},  
W49N appears to be at a similar evolutionary stage to the supernebula in NGC 5253. 
W49N has a luminosity of $7 \times 10^6 $~\lsun\citep{1996MNRAS.281..294B}, 100 times less
 that of the supernebula.
Yet W49N has a larger, perhaps significantly larger, molecular gas mass,
 $\sim 1.2\times 10^5 $~\msun\ \citep{2013ApJ...779..121G}. 
 Even including the cloud to the southwest of the supernebula, of similar mass to
 the W49N cloud \citep{SMC17},  the supernebula region
  is $\sim$50 times more efficient than W49N at forming stars in terms of $L_{IR}/M_{H2}$. 
  
 An alternative to high star formation efficiency is that gas has already dispersed from the
 young cluster. Based on the CO linewidth, dispersal of gas from the 
 cluster 
 does not appear to be occurring at present. 
  Given the young age of the cluster, any gas that has previously been dispersed
  by the cluster  cannot be far away. Gas expelled at 
 $\sim$20 \kms, the overall velocity dispersion within the region \citep{SMC17}
 would still be within $\sim$20-60 pc of the cluster given its age; 
 the clouds will still lie  within Figure 1. 
 
 The continued presence of dense molecular gas within the supernebula cluster and its 
 relative quiescence suggests that negative star formation feedback effects are minimal on the 
 molecular gas at this stage of cluster evolution of the supernebula, 
 perhaps because the gas is in the form of compact clumps or cores. The high
 density indicated by the CO(3--2) emission suggests that 
 Cloud D1 may still be forming stars.  
  The CO(3-2) emission may arise in dense hot molecular cores around young stellar objects,
  which would be   consistent with the warm, 
$T_K \ga 300K$ temperature suggested  by lower resolution CO line ratios \citep{2015Natur.519..331T},
and the optically thin CO(3--2) emission reported here. 
 For this temperature, and given that the CO cores are embedded 
 in an HII region of density $n_H =3.5 \times 10^4~\rm cm^{-3}$ \citep{2004ApJ...602L..85T}, 
 the Bonnor-Ebert mass of Cloud D1  is $ M_{BE }\sim$ 13 \msun. 
  If stars are still forming within the core of the cluster embedded in Cloud D1, they will be massive.

\section{Conclusions} \label{sec:conclusions}

We present  ALMA observations of CO(3--2) and $^{13}$CO(3--2)
at 0\farcs3 resolution of the region surrounding the 
supernebula  in NGC 5253.
We identify an unusual  cloud, ``Cloud D1", that is precisely coincident with the supernebula/
embedded cluster in 
space and velocity.
Cloud D1 has a radius of r$\sim$2.8 pc, nearly the same size as the supernebula. 
Based on\\
1)  spatial coincidence to $<0.6$~pc in projection with the supernebula, \\
2)  velocity coincidence to within $2$-$3$~\kms\ with mid-IR nebular and radio recombination lines 
from the HII region and \\
3) the fact that the cloud
is  optically thin, probably because it is hot \\
we conclude that 
the molecular cloud is mixed in with the 
super star cluster, which contains $\sim$1500 O stars, and the compact HII region. 

The CO linewidth of 21.7 \kms\ indicates that the CO gas in Cloud D1 is relatively undisturbed
in spite of its location within this dense cluster. 
The linewidth indicates a  dynamical mass of $M_{dyn}=2.5\times 10^5$~\msun. This gives
$M/L \sim 5 \times 10^{-4}~\rm M_\odot/L_\odot$ for the cluster, implying 
a top-heavy IMF, with lower cutoff of $\ga$1--2~\msun, and
more if there is photon leakage beyond the supernebula. 

A CO(3--2)/$^{13}$CO ratio of 
$\sim$50 indicates the the emission from Cloud D1 is bright because the gas is
 optically thin and not because there is a high gas column.  
Estimates of gas mass based on this line are  uncertain, 
but suggest that the molecular gas mass  is $<$60,000~\msun. 
The star formation efficiency appears
to be $\ga$50 times that of W49N, the closest Galactic analogue. The 
smoothness of the CO line profile and its near Gaussianity
suggests  Cloud D1 consists of many, up to hundreds, of
molecular clumps or cores. Given the high ambient pressure and temperature within Cloud D1, if
the cloud is indeed  hot as previously estimated \citep{2015Natur.519..331T}
only massive stars $\ga$13~\msun\ can form at present. 
We propose that Cloud D1 is composed of many 
hot molecular clumps or cores orbiting within the cluster potential with the stars of the 
super star cluster, and may yet be capable of forming
stars.

\acknowledgments

This paper makes use of the following ALMA data: 
ADS/JAO.ALMA\# 2012.1.00125.S. ALMA is a partnership of ESO (representing its member states), 
NSF (USA) and NINS (Japan), together with NRC (Canada) and NSC and ASIAA (Taiwan), 
in cooperation with the Republic of Chile. The Joint ALMA Observatory is operated by ESO, 
AUI/NRAO and NAOJ. The National Radio Astronomy Observatory is a facility of the 
National Science Foundation (NSF) operated under cooperative agreement by Associated Universities, Inc. 
SMC acknowledges the support of an NRAO Student Observing Support Grant. 
The authors thank  Adam Ginsburg, Christian Henkel, Leslie Hunt, Pavel Kroupa, Phil Myers,
Nick Scoville,  and an anonymous referee for helpful discussions and comments. 
JLT acknowledges additional support from NSF grant AST 1515570, and from a COR grant from the UCLA Academic Senate.

%

\vspace{5mm}
\facilities{ALMA}


\software{Starburst99 \citep{1999ApJS..123....3L},  CASA, AIPS, SAOImage DS9 }


\clearpage

\begin{thebibliography}{}

\bibitem[Alonso-Herrero et al.(2004)]{2004ApJ...612..222A} 
  Alonso-Herrero, A., Takagi, T., Baker, A.~J., et al.\ 2004, \apj, 612, 222
  \bibitem[Banerjee et al.(2012)]{2012MNRAS.426.1416B} 
  Banerjee, S., Kroupa, P., \& Oh, S.\ 2012, \mnras, 426, 1416
\bibitem[Beck(2008)]{2008A&A...489..567B} Beck, S.~C.\ 2008, \aap, 489, 567 
\bibitem[Beck et al.(2012)]{2012ApJ...755...59B} 
Beck, S.~C., Lacy, J.~H., Turner, J.~L., et al.\ 2012, \apj, 755, 59 
\bibitem[Beck et al.(1996)]{1996ApJ...457..610B} 
Beck, S.~C., Turner, J.~L., Ho, P.~T.~P., Lacy, J.~H., \& Kelly, D.~M.\ 1996, \apj, 457, 610 
\bibitem[Bendo et al.(2017)]{Bendo17}
Bendo, G. J., Miura, R. E., Espada, D., Nakanishi, K., Beswick, R. J., D'Cruze, M. J., Dickinson, C., \& 
Fuller, G. A., \mnras, submitted.
\bibitem[Bisbas et al.(2015)]{2015MNRAS.454.2828B} 
Bisbas, T.~G., Haworth, T.~J., Barlow, M.~J., et al.\ 2015, \mnras, 454, 2828 
\bibitem[Bolatto et al.(2013)]{2013Natur.499..450B} 
Bolatto, A.~D., Warren, S.~R., Leroy, A.~K., et al.\ 2013, \nat, 499, 450 
\bibitem[Buckley \& Ward-Thompson(1996)]{1996MNRAS.281..294B} 
Buckley, H.~D., \& Ward-Thompson, D.\ 1996, \mnras, 281, 294 
\bibitem[Burbidge \& Burbidge(1962)]{1962ApJ...135..694B} 
Burbidge, E.~M., \& Burbidge, G.~R.\ 1962, \apj, 135, 694 
\bibitem[Caldwell \& Phillips(1989)]{1989ApJ...338..789C} 
Caldwell, N., \& Phillips, M.~M.\ 1989, \apj, 338, 789 
\bibitem[Calzetti et al.(1997)]{1997AJ....114.1834C}
 Calzetti, D., Meurer, G.~R., Bohlin, R.~C., et al.\ 1997, \aj, 114, 1834 
 \bibitem[Calzetti et al.(2015)]{2015ApJ...811...75C} 
 Calzetti, D., Johnson, K.~E., Adamo, A., et al.\ 2015, \apj, 811, 75 
 \bibitem[Consiglio et al.(2017)]{SMC17}
 Consiglio et al., submitted. 
 \bibitem[Cresci et al.(2005)]{2005A&A...433..447C} 
 Cresci, G., Vanzi, L., \& Sauvage, M.\ 2005, \aap, 433, 447 
 \bibitem[Crowther et al.(2010)]{2010MNRAS.408..731C} 
 Crowther, P.~A., Schnurr, O., Hirschi, R., et al.\ 2010, \mnras, 408, 731 
 \bibitem[de Grijs et al.(2013)]{2013MNRAS.431.2917D} 
 de Grijs, R., Anders, P., Zackrisson, E., \& {\"O}stlin, G.\ 2013, \mnras, 431, 2917
 \bibitem[De Pree et al.(2000)]{2000ApJ...540..308D} 
 De Pree, C.~G., Wilner, D.~J., Goss, W.~M., Welch, W.~J., \& McGrath, E.\ 2000, \apj, 540, 308 
 \bibitem[Dreher et al.(1984)]{1984ApJ...283..632D} 
 Dreher, J.~W., Johnston, K.~J., Welch, W.~J., \& Walker, R.~C.\ 1984, \apj, 283, 632 
 \bibitem[Galv{\'a}n-Madrid et al.(2013)]{2013ApJ...779..121G} 
 Galv{\'a}n-Madrid, R., Liu, H.~B., Zhang, Z.-Y., et al.\ 2013, \apj, 779, 121 
 \bibitem[Gorjian(1996)]{1996AJ....112.1886G} 
 Gorjian, V.\ 1996, \aj, 112, 1886 
 \bibitem[Gorjian et al.(2001)]{2001ApJ...554L..29G} 
 Gorjian, V., Turner, J.~L., \& Beck, S.~C.\ 2001, \apjl, 554, L29 
\bibitem[Harris et al.(2004)]{2004ApJ...603..503H} 
Harris, J., Calzetti, D., Gallagher, J.~S., III, Smith, D.~A., \& Conselice, C.~J.\ 2004, \apj, 603, 503 
\bibitem[Heikkil{\"a} et al.(1999)]{1999AA...344..817H} 
Heikkil{\"a}, A., Johansson, L.~E.~B., \& Olofsson, H.\ 1999, \aap, 344, 817 
\bibitem[Hollenbach \& Tielens(1999)]{HT99} Hollenbach, D.~J., \& Tielens, A.~G.~G.~M.\ 1999, Reviews of Modern Physics, 71, 173 
\bibitem[Hunt et al.(2005)]{2005A&A...434..849H} 
 Hunt, L., Bianchi, S., \& Maiolino, R.\ 2005, \aap, 434, 849
 \bibitem[Inoue(2001)]{2001AJ....122.1788I} 
 Inoue, A.~K.\ 2001, \aj, 122, 1788 
 \bibitem[Kleinmann \& Low(1970)]{1970ApJ...159L.165K} 
 Kleinmann, D.~E., \& Low, F.~J.\ 1970, \apjl, 159, L165 
\bibitem[Kobulnicky \& Skillman(2008)]{2008AJ....135..527K} 
Kobulnicky, H.~A., \& Skillman, E.~D.\ 2008, \aj, 135, 527 
\bibitem[Kobulnicky et al.(1997)]{1997ApJ...477..679K} Kobulnicky, H.~A., 
Skillman, E.~D., Roy, J.-R., Walsh, J.~R., 
\& Rosa, M.~R.\ 1997, \apj, 477, 679
\bibitem[Kroupa \& Boily(2002)]{2002MNRAS.336.1188K} 
Kroupa, P., \& Boily, C.~M.\ 2002, \mnras, 336, 1188 
\bibitem[Kroupa et al.(2013)]{2013pss5.book..115K} 
Kroupa, P., Weidner, C., Pflamm-Altenburg, J., et al.\ 2013, 
Planets, Stars and Stellar Systems.~Volume 5: Galactic Structure and Stellar Populations, 5, 115 
\bibitem[Lada et al.(2010)]{2010ApJ...724..687L} 
Lada, C.~J., Lombardi, M., \& Alves, J.~F.\ 2010, \apj, 724, 687 
\bibitem[Leitherer et al.(2014)]{2014ApJS..212...14L} 
Leitherer, C., Ekstr{\"o}m, S., Meynet, G., et al.\ 2014, \apjs, 212, 14 
\bibitem[Leitherer et al.(1999)]{1999ApJS..123....3L} 
Leitherer, C., Schaerer, D., Goldader, J.~D., et al.\ 1999, \apjs, 123, 3 
 \bibitem[L{\'o}pez-S{\'a}nchez et al.(2007)]{2007ApJ...656..168L} L{\'o}pez-S{\'a}nchez, {\'A}.~R., Esteban, C., Garc{\'{\i}}a-Rojas, J., Peimbert, M., \& Rodr{\'{\i}}guez, M.\ 2007, \apj, 656, 168 
 \bibitem[Lundgren(2013)]{L2013}
 Lundgren, A. ALMA Cycle 2 Technical Handbook Version 1.1, ALMA
 \bibitem[Martin(1998)]{1998ApJ...506..222M} 
Martin, C.~L.\ 1998, \apj, 506, 222 
\bibitem[Mart{\'{\i}}n-Hern{\'a}ndez et al.(2005)]{2005A&A...429..449M} 
Mart{\'{\i}}n-Hern{\'a}ndez, N.~L., Schaerer, D., \& Sauvage, M.\ 2005, \aap, 429, 449 
\bibitem[Meier et al.(2002)]{2002AJ....124..877M} 
Meier, D.~S., Turner, J.~L., \& Beck, S.~C.\ 2002, \aj, 124, 877 
\bibitem[Meurer et al.(1995)]{1995AJ....110.2665M} 
Meurer, G.~R., Heckman, T.~M., Leitherer, C., et al.\ 1995, \aj, 110, 2665 
\bibitem[Miura et al.(2015)]{2015PASJ...67L...1M} 
Miura, R.~E., Espada, D., Sugai, H., Nakanishi, K., \& Hirota, A.\ 2015, \pasj, 67, L1 
\bibitem[Murray et al.(2010)]{2010ApJ...709..191M} 
Murray, N., Quataert, E., \& Thompson, T.~A.\ 2010, \apj, 709, 191 
\bibitem[Myers(1985)]{1985prpl.conf...81M} 
Myers, P.~C.\ 1985, Protostars and Planets II, 81 
\bibitem[Oey \& Clarke(2005)]{2005ApJ...620L..43O} 
Oey, M.~S., \& Clarke, C.~J.\ 2005, \apjl, 620, L43 
\bibitem[Rodr{\'{\i}}guez-Rico et al.(2007)]{2007ApJ...670..295R} 
Rodr{\'{\i}}guez-Rico, C.~A., Goss, W.~M., Turner, J.~L., \& G{\'o}mez, Y.\ 2007, \apj, 670, 295 
\bibitem[Schaerer et al.(1997)]{1997ApJ...481L..75S} 
Schaerer, D., Contini, T., Kunth, D., \& Meynet, G.\ 1997, \apjl, 481, L75 
\bibitem[Sakamoto et al.(2006)]{2006ApJ...636..685S} 
Sakamoto, K., Ho, P.~T.~P., Iono, D., et al.\ 2006, \apj, 636, 685 
\bibitem[Schilke et al.(1997)]{1997ApJS..108..301S} 
Schilke, P., Groesbeck, T.~D., Blake, G.~A., Phillips, \& T.~G.\ 1997, \apjs, 108, 301 
\bibitem[Silich \& Tenorio-Tagle(2017)]{2017MNRAS.465.1375S} 
Silich, S., \& Tenorio-Tagle, G.\ 2017, \mnras, 465, 1375 
\bibitem[Smith et al.(2016)]{2016ApJ...823...38S} 
Smith, L.~J., Crowther, P.~A., Calzetti, D., \& Sidoli, F.\ 2016, \apj, 823, 38 
\bibitem[Stanway et al.(2016)]{2016MNRAS.456..485S} 
Stanway, E.~R., Eldridge, J.~J., \& Becker, G.~D.\ 2016, \mnras, 456, 485
\bibitem[Tan et al.(2014)]{2014prpl.conf..149T} 
Tan, J.~C., Beltr{\'a}n, M.~T., Caselli, P., et al.\ 2014, Protostars and Planets VI, 149 
\bibitem[Tenorio-Tagle et al.(2015)]{2015ApJ...814L...8T} 
Tenorio-Tagle, G., Mu{\~n}oz-Tu{\~n}{\'o}n, C., Silich, S., \& Cassisi, S.\ 2015, \apjl, 814, L8 
\bibitem[Tremonti et al.(2001)]{2001ApJ...555..322T} 
Tremonti, C.~A., Calzetti, D., Leitherer, C., \& Heckman, T.~M.\ 2001, \apj, 555, 322 
\bibitem[Turner \& Beck(2004)]{2004ApJ...602L..85T} 
Turner, J.~L., \& Beck, S.~C.\ 2004, \apjl, 602, L85 
\bibitem[Turner et al.(2003)]{2003Natur.423..621T} 
Turner, J.~L., Beck, S.~C., Crosthwaite, L.~P., et al.\ 2003, \nat, 423, 621 
\bibitem[Turner et al.(2000)]{2000ApJ...532L.109T} 
Turner, J.~L., Beck, S.~C., \& Ho, P.~T.~P.\ 2000, \apjl, 532, L109 
\bibitem[Turner et al.(1997)]{1997ApJ...474L..11T} 
Turner, J.~L., Beck, S.~C., \& Hurt, R.~L.\ 1997, \apjl, 474, L11 
\bibitem[Turner et al.(1998)]{1998AJ....116.1212T} 
Turner, J.~L., Ho, P.~T.~P., \& Beck, S.~C.\ 1998, \aj, 116, 1212 
\bibitem[Turner et al.(2015)]{2015Natur.519..331T}
Turner, J.L., Beck, S.C., Benford, D.J., et al. \ 2015, Nature, 519, 7543
\bibitem[Vanzi \& Sauvage(2004)]{2004A&A...415..509V} 
Vanzi, L., \& Sauvage, M.\ 2004, \aap, 415, 509 
\bibitem[Walsh \& Roy(1989)]{1989MNRAS.239..297W} 
Walsh, J.~R., \& Roy, J.-R.\ 1989, \mnras, 239, 297 
\bibitem[Weidner \& Kroupa(2004)]{2004MNRAS.348..187W} 
Weidner, C., \& Kroupa, P.\ 2004, \mnras, 348, 187 



\end{thebibliography}
\end{document}